# Efficient similar waveform search using short binary codes obtained through a deep hashing technique

Makoto Naoi[1] and Shiro Hirano[2]

[1]Research Center for Earthquake Hazards, Disaster Prevention Research Institute, Kyoto University, Gokasyo, Uji, Kyoto 611-0011, Japan.

[2]College of Science and Engineering, Ritsumeikan University, 1-1-1 Noji Higashi, Kusatsu, 525-8577, Japan.

**Abbreviated title:** Similar waveform search via deep hashing

**Corresponding author**: Makoto Naoi (E-mail: naoi.makoto.4z@kyoto-u.ac.jp)

## SUMMARY

A similar waveform search plays a crucial role in seismology for detecting seismic events, such as small earthquakes and low-frequency events. However, the high computational costs associated with waveform cross-correlation calculations represent bottlenecks during the analysis of long, continuous records obtained from numerous stations. In this study, we developed a deep-learning network to obtain 64-bit hash codes containing information on seismic waveforms. Using this network, we performed a similar waveform search for ~35 million moving windows developed for the 30 min waveforms recorded continuously at 10 MHz sampling rates using 16 acoustic emission transducers during a laboratory hydraulic fracturing experiment. The sampling points of each channel corresponded to those of the 5.8-year records obtained from typical seismic observations at 100 Hz sampling rates. Of the 35 million windows, we searched for windows with small average Hamming distances among the hash codes of 16 channel waveforms against template hash codes of 6057 events that were catalogued using conventional autoprocessing techniques. The calculation of average Hamming distances is 1000 times faster than that of the corresponding network correlation. This hashing-based template matching enabled the detection of 23,462 additional events. We also demonstrated the feasibility of the hashing-based autocorrelation analysis, where similar event pairs were extracted without templates, by calculating the average Hamming distances for all possible pairs of the ~35 million windows. This calculation required only 15.5 h under 120 thread parallelisation. This deep hashing approach significantly reduced the required memory compared with locality-sensitive hashing approaches based on random permutations, enabling similar waveform searching on a large-scale dataset.

# 1. INTRODUCTION

A similar waveform search is a powerful tool for identifying repeating earthquakes (Igarashi et al. 2003) and detecting seismic events with small amplitudes (Gibbons & Ringdal 2006). For example, matched filter analysis or template matching, wherein windows exhibiting high correlation coefficients with template waveforms (Peng & Zhao 2009) are extracted from continuous records, often increases the number of available events ten-fold, demonstrating their activities at higher spatiotemporal resolutions. This technique facilitates detailed investigations of seismicity, for example, preceding large earthquakes (Bouchon 2011; Doi & Kawakata 2012; 2013; Kato et al. 2012). Furthermore, autocorrelation analysis, wherein windows containing similar waveforms are identified without templates based on cross-correlation coefficients, helps detect low-frequency earthquakes (Brown et al. 2008; Aguiar & Beroza 2014).

Previous studies have repeatedly demonstrated the effectiveness of analyses based on cross-correlations. However, high computational costs hinder the applications of these methods to large-scale datasets. For example, Ross et al. (2019) performed template matching for 10-year continuous records of ~500 stations with 0.28 million template events. They required significant computer resources and used an array of 200 NVIDIA P100 graphics processing units (GPUs). Although previous studies have attempted to reduce these calculation costs (Skoumal et al. 2016; Senobari et al. 2019), cross-correlation calculations among numerous waveforms are not essentially scalable. Moreover, recent approaches employing synthetic template waveforms (Chamberlain & Townend 2018; Ide 2021) may also aggravate the requirements of efficient similarity searches in the event detection problem.

In image/audio recognition problems, approximate nearest-neighbour search algorithms (Arya et al. 1998) have been used to efficiently extract similar objects from content. A few of these algorithms have been applied to seismic records (Yoon et al. 2015; Tibi et al. 2017). For example, an approach based on locally sensitive hashing (LSH), that is, hash functions mapping similar objects to feature vectors with small distances in the mapped hash space, has been demonstrated to be effective. In the ideal case of Hashing approaches, an object with a specific hash code can be identified by simply looking up the table that has the keys of the hash code and the values of the target objects (or their identifiers) with a computing cost of only $O(1)$. In the LSH approach, objects with similar hash codes are probabilistically searched, enabling much faster similar object searching than the method with distance/similarity calculations among all object pairs. Yoon et al. (2015) developed the fingerprint and similarity thresholding (FAST) algorithm to perform a similar waveform search using a LSH hashing method for seismic records. In this method, the authors first converted a spectrogram of a seismic waveform into a fingerprint (4096-bit binary codes in the case of Yoon et al. 2015), which was followed by hashing for similar waveform grouping. This approach achieved a computational cost of $O(N^{1.36})$ for a similar waveform search (including a $O(N)$ cost to obtain the hash codes for $N$ objects) among all the possible pairs of waveforms extracted from a continuous record using overlapping moving windows. This approach was much more scalable than the corresponding analysis based on cross-correlation calculations (i.e. autocorrelation analysis; Brown et al. 2008), which require the computational costs of $O(N^2)$.

In the classical LSH approach used by Yoon et al. (2015), hash functions such as the random permutation of vector components (Min-Hash; Broder et al. 2000), which are independent of the data structure and can be used for various objectives, are adopted. These simple functions require multiple hashings for one object to ensure accuracy, and similar objects are identified as those for which the number of resultant codes exceeds a predetermined threshold. However, this approach requires a large amount of computer memory to hold numerous hash tables for efficient calculations, which may hinder its application to large datasets.

This problem can be addressed using a hash function that generates small codes with rich information regarding original waveforms. Weiss et al. (2008) suggested such an approach called 'spectral hashing' by incorporating structures of the applied dataset. Notably, such data-dependent hashing generates short (e.g. 64 bit) codes that enable faster and memory-saving nearest-neighbour searches. Recently, more efficient hashing approaches have been developed using deep learning techniques that can learn complicated nonlinear mappings (Lu et al. 2017). Recent advances in deep hashing techniques have been reviewed by Singh & Gupta (2022) and Luo et al. (2020).

In this study, we applied a deep hashing technique to continuous waveforms recorded by acoustic emission (AE) transducers in a laboratory for a fast, memory-saving similarity search. We analysed the waveform records obtained from hydraulic fracturing experiments conducted by Tanaka et al. (2021). A deep hashing network was trained to map a 1024-sample waveform to a single 64-bit binary code based on the AE catalogue developed by Tanaka et al. (2021) using conventional autoprocessing algorithms. We searched for similar events in 16-channel continuous waveforms recorded at a 10 MHz sampling rate using the obtained model and resultant hash codes during a ~30 min experiment to detect additional AE events. The number of sampling points at each channel approached $1.8 \times 10^{10}$, corresponding to a 5.8-year continuous record at a 100 Hz sampling rate, which is typically adopted in surface seismic observations.

## 2. Method

### *2.1 Network architecture and loss functions used to obtain hash functions*

Inspired by Huang et al. (2017), who developed a hashing model for an image retrieval problem, we developed a deep-learning network to hash seismic waveforms. In the foregoing study, to train the network, the authors prepared numerous triplets of images consisting of an anchor image, a positive (similar) sample obtained by rotating the anchor image, and a negative (dissimilar) sample, which was a different image from that of the anchor. They trained a deep hashing network using the triplets to obtain real-valued (0–1) embedding vectors from an image by minimising the weighted sum of the triplet loss (Weinberger & Saul 2009), quantisation loss, and entropy loss (Huang et al. 2017). These losses were calculated from the resulting embedding vectors. Following this, hash codes were obtained to retrieve similar objects by rounding each embedding vector component to a binary value of either 0 or 1. In this study, we modified this framework to analyse seismic waveforms. Table 1 summarises the network used to obtain a 1 × 64 real-valued embedding vector from a waveform, and Fig. 1 presents the network used to train the network presented in Table 1.

Table. 1 Architecture of the deep learning network used to obtain a real-valued embedding vector from a waveform. The kernel size of one-dimensional convolution (Conv1D) is three. The convolution was conducted with zero padding to obtain same-size output tensors as the input tensors. The network in Fig. 1 incorporates this network for training.

|  | Input size | output size | No. of channels | activation |
|---|---|---|---|---|
| Conv1D + Conv1D + Max. pooling | 1024 x 1 | 512 x 32 | 32 | relu |
| Conv1D + Conv1D + Max. pooling | 512 x 32 | 256 x 32 | 32 | relu |
| Conv1D + Conv1D + Max. pooling | 256 x 32 | 128 x 32 | 32 | relu |
| Conv1D + Conv1D + Max. pooling | 128 x 32 | 64 x 32 | 32 | relu |
| Conv1D + Conv1D + Max. pooling | 64 x 32 | 32 x 32 | 32 | relu |
| Conv1D + Conv1D + Max. pooling | 32 x 32 | 16 x 32 | 32 | relu |
| Conv1D + Conv1D + Max. pooling | 16 x 32 | 8 x 32 | 32 | relu |
| Flatten | 8 x 32 | 1 x 256 | 1 |  |
| Dense | 1 x 256 | 1 x 64 | 1 | sigmoid |

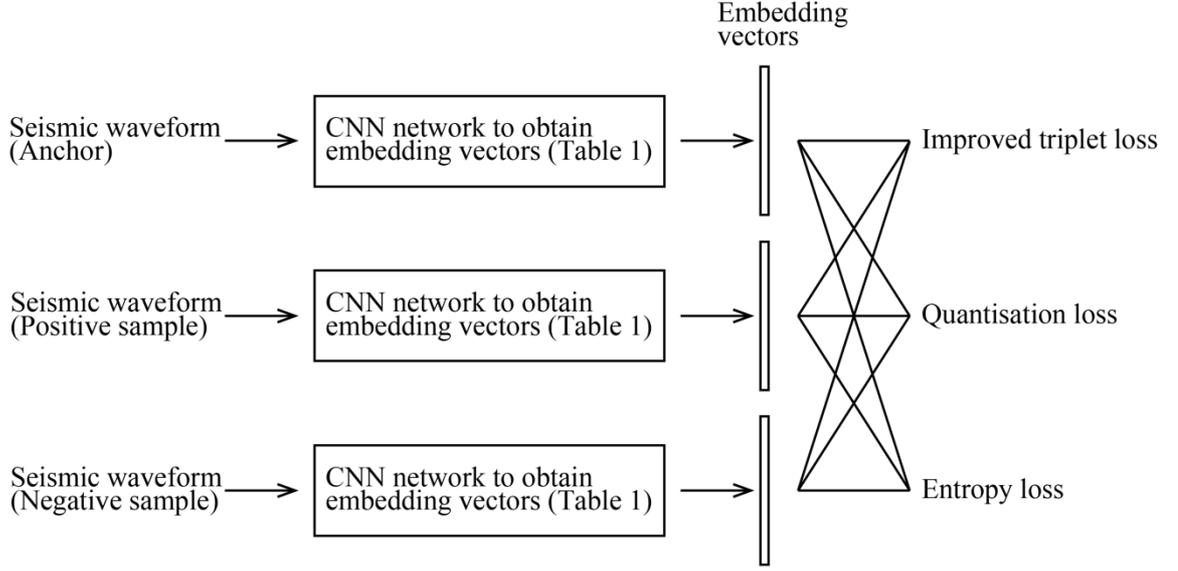

Fig. 1. Deep-learning network used to train the network depicted in Table 1. Anchor, positive, and negative samples are input to the same network (i.e. it has the same weight and structure shown in Table 1) to obtain respective embedding vectors.

The triplet loss (Weinberger & Saul 2009) is defined as max(0, $D_n^+$–$D_n^-$+$\theta_1$), where $D_n^+$ denotes the distance between a pair of embedding vectors for the anchor and positive samples in the nth triplet, and $D_n^-$ represents the distance of the anchor-negative pair. Each embedding vector for the distance calculation is obtained through the exactly same network with same weights. $\theta_1$ denotes a margin promoted between positive and negative pairs. By minimising this loss function, a deep hashing network learns to generate embedding vectors satisfying $D_n^+ + \theta_1 < D_n^-$ for each triplet. Instead of the original triplet loss, we used the following improved triplet loss (Cheng et al. 2016)

$$L_T = \frac{1}{2N}\sum_{n=1}^{N}[max\{0, D_n^+ - D_n^- + \theta_1\} + max\{0, D_n^+ - \theta_2\}], \qquad (1)$$

for constraining, so that the distances of anchor-positive pairs were smaller than $\theta_2$, where $N$ denotes the number of triplets in the training data. We used the Euclidean distance as the distance metric $D$ between the embedding vectors. Although Huang et al. (2017) calculated the triplet loss based on the square of the Euclidean distance after normalisation, we avoided that treatment partially because a direct comparison of $\theta_1$ and $\theta_2$ with the Euclidean distance between embedding vectors is feasible and because the upper bound of $|D^+–D^-|$ can be determined based on the triangular inequality.

According to Huang et al. (2017), we used the quantisation loss $L_Q$ defined as

$$L_Q = \frac{1}{2N}\sum_{n=1}^{N}\|F(p_n) - b_n\|, \qquad (2)$$

and the entropy loss $L_E$ defined as

$$L_E = \frac{1}{2M}\sum_{m=1}^{M}(\mu_m - 0.5), where\ \mu_m = \frac{1}{N}\sum_{n=1}^{N} b_n(m), \qquad (3)$$

where $M$ denotes the dimension of the embedding vector (64), $p_n$ denotes the nth waveform sample, $F$ denotes the mapping function corresponding to the deep hashing network (the network listed in Table 1). $F(p_n)$ ranges between 0 and 1 due to the Sigmoid activation function of the final layer (Table 1). $b_n$ denotes the 64-bit hash code corresponding to the rounded $F(p_n)$ with 0.5 as the threshold, and $b_n(m)$ in Equation (3) denotes $m$th components consisting of the binary vector. The quantisation loss $L_Q$ makes real values composing the embedding vectors approach zero or one, thereby reducing the quantisation error (i.e. the difference between the hash codes and embedding vectors). The entropy loss $L_E$ balances the appearing rates of ones and zeros in the hash codes to store information efficiently (Huang et al. 2017). During the training process, we minimised the weighted sum of the three loss functions:

$$L = \alpha L_T + \beta L_Q + \gamma L_E, \qquad (4)$$

where $\alpha$, $\beta$, and $\gamma$ denote the weight of each function.

## *2.2 Training the deep hashing network using laboratory-AE waveforms*

Huang et al. (2017) created a positive sample by rotating an anchor image and randomly selected another image as a negative sample to train a hash function using a deep learning network similar to that depicted in Fig. 1. Inspired by this approach, we prepared an anchor and a positive sample by adding a time shift and noise to the same original waveform and prepared negative samples using the same procedure based on another original waveform. The procedure is described in detail in the following section.

In this study, we analysed the waveform records of AE transducers obtained by Tanaka et al. (2021) in laboratory hydraulic fracturing experiments. The experiments were conducted using samples of Kurokami-jima granite with dimensions of 65 × 65 × 130 mm$^3$ under a uniaxial loading of 5 MPa. A thermosetting acrylic resin was injected into a packer set in a 6 mm borehole to induce hydraulic fracturing. Tanaka et al. (2021) recorded signals using 16 high-sensitivity broadband transducers (M304A, Fuji Ceramics Corp.) and eight traditional AE transducers (PICO: Physical Acoustics Co.). The waveforms were continuously recorded at a 10 MHz sampling rate and stored as 16-bit integer variables using a 14-bit analogue–digital converter. An analogue 0.02–3.00 MHz bandpass filter was applied only for M304A before the analogue–digital conversion.

Tanaka et al. (2021) conducted fracturing experiments on 10 specimens. Among these, in this study, we analysed the continuous waveforms obtained from an experiment on specimen KJG1810. For this experiment, we searched for similar AE events in the records using a deep-hash model (Table 1) trained using the dataset based on the AE catalogue developed by Tanaka et al. (2021). In their cataloguing procedure, the authors detected AE events using the ratio between short- and long-term averages (Allen 1978) and estimated *P*-arrivals using the algorithm proposed by Tanakami & Kitagawa (1988) to determine their hypocentres. In total, 6057 events were listed in their AE catalogue, and we used the corresponding waveform records of the 16 M304A transducers for training. We used the same number of noise waves extracted from the waveform record before the start of AE activity. The total number of training triplets was 193,824 (6057×16×2). All waveforms were used after the application of a 50–300 Hz bandpass filter. Although Tanaka et al. (2021) relocated the hypocentres using the double-difference method (Waldhauser & Ellsworth 2000), we used unrelocated results because the relocation process reduced the number of available events.

Generally, a part of the dataset prepared for training is utilised as validation and test data in the training of a deep learning network to check for overfitting and assess the final performance. In the present study, this corresponds to the preparation of the validation and test data based on the KJG1810 catalogue. However, we considered that all template waveforms in the catalogue should be included in the training data when using the model to solve template matching problems (Section 4), similar to the case of a music-searching system (e.g. Arcas et al. 2017). Therefore, we prepared the validation and test data based on the AE catalogue of the KJG1811 experiment, which was conducted under the same experimental conditions as those for KJG1810. For the KJG1811 experiment, Tanaka et al. (2022) developed an AE catalogue consisting of 3864 events. We created triplets from the corresponding waveforms and randomly separated the results using an equal ratio of the validation and test data, resulting in 61,824 validation/test triplets, including noise triplets (3864/2 × 16 × 2).

We extracted the waveforms of the catalogued events recorded by the 16 M304A transducers using a 1024 sample window to create an anchor waveform in each triplet. The window began with 150 samples before the minimum theoretical arrival time among the 16 channels, and random perturbation within +/– 300 samples was added. After normalisation with the absolute maximum amplitude, we added uniform white noise, whose absolute maximum amplitudes were randomly determined within +/– 0.5, to the cutout waveform. Positive and negative samples were then created by applying the same procedure to the same and different original waveforms. We also prepared triplets based on the noise waveforms extracted from the period before the initiation of AE activity. In Section 3.2, we describe the contribution of these noise triplets to performance improvements.

For training based on the triplet loss, semi-hard negative or hard negative samples are useful for efficient training (Kaya & Bilge 2019). For positive and negative samples at distances of $D^+$ and $D^-$ from the anchor sample, respectively, a semi-hard negative is defined as the sample with $D^-$ satisfying

$$D^+ < D^- < D^+ + \theta_l, \quad (5)$$

and a hard negative is defined as a sample with $D^-$ satisfying

$$D^- < D^+. \quad (6)$$

Semi-hard and hard negative samples correspond to closely misjudged or misjudged samples by the hash function. Hence, training using these samples efficiently increases the discrimination capability of the model. Conversely, the easy negative sample satisfying

$$D^- > D^+ + \theta_1 \quad (7)$$

corresponds to one whose embedding vector significantly differs from the anchor, and its contribution to training is limited (Kaya & Bilge 2019). For efficient training, we preferably selected semi-hard and hard negative samples during the training process, as will be described below. We prepared a negative sample for each anchor based on a randomly selected waveform that differed from that of the anchor at the beginning of the training process. After 20 training epochs, we re-created the anchor and positive samples based on the same original waveforms by adding different time shifts and white noise. This is a kind of augmentation process to mitigate overfitting. We also selected 10 negative samples for each anchor to generate 10 triplets. Negative samples were randomly selected within each of the following (in the order of priority), semi-hard negative, hard negative, and easy negative, determined based on the embedding vectors generated by the deep-hash network at the epoch and Euclidean distances for all possible pairs among the re-created positive samples. The training during each epoch was conducted using a dataset with one triplet for each anchor. Hence, 10 datasets were generated with different negative samples in the re-creation process. Ten epochs were considered for each dataset training. Subsequently (i.e. a total of 120 epochs were completed), the re-creation and training processes were repeated until the number of re-creations reached 10 times. In total, 1020 epochs were used in the training process.

We used margin parameters of $\theta_1 = \sqrt{3}$ and $\theta_2 = 1$ (eq. (1)), corresponding to Hamming distances of $d_h = 3$ and 1, respectively. Note that the Hamming distance is defined as the number of different bits between two binary codes (which correlates with the Euclidean distance used in the training process) and is frequently used as a distance metric for binary codes. When a small $\theta_1$ is adopted, the obtained hash codes occupy only a small portion of the 64-bit binary space, reducing the resolution to distinguish waveforms. We tested several values and adopted a value of $\sqrt{3}$ so that the maximum Hamming distance between the hash codes obtained for waveforms of the 6057 events, which was used in training, was ~50. The Hamming distances between pairs increased with the value of $\theta_1$, although we used $\theta_2 = 1$ to constrain the anchor-positive pairs to have the same code. Although we did not perform exhaustive parameter searches, these parameters functioned well in the subsequent analyses. For the weight parameters in Eq. (4), we used the values of $\alpha = 100$ and $\beta = \gamma = 1$ because a large value of $\beta$ compared to that of $\alpha$ prevented learning progress during training. A larger quantisation loss weight likely results in embedding vectors composed of only zeros and ones, which significantly reduces the amount of information.

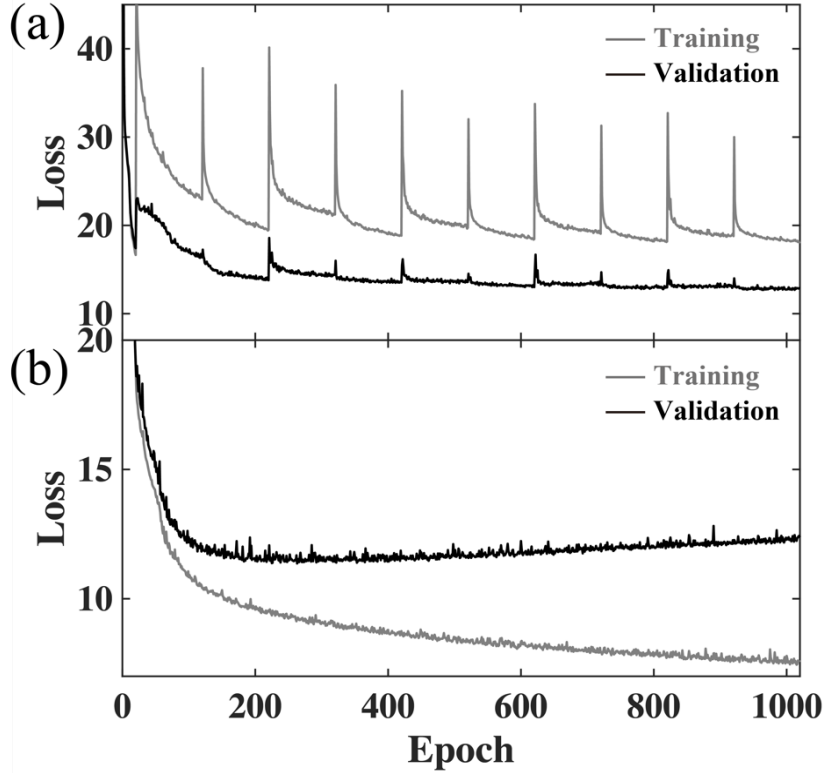

Fig. 2. Learning curves obtained during the training of the network in Fig. 1. Grey lines represent the loss in the training data, and black lines represent the loss in the validation data. (a) Results obtained for training with the re-creation of the training dataset. (b) Results for training without recreation.

Fig. 2(a) presents the learning curve obtained during training. The 1020-epoch training took 4.4 hours using two NVIDIA A40 GPU. After the first 20 epochs, the loss of the training data exceeded that of the validation data because semi-hard and hard negative samples were preferably selected. When the training data were re-created after every 100 epochs, the loss values suddenly increased, particularly for the training data. For comparison, Fig. 2(b) presents the loss history for the case without training data re-creation (i.e. the same training data were used throughout all epochs). In this case, deviation of the loss values for training and validation data begin to increase by ~70 epochs, indicating the progress of overfitting. The validation loss reached a minimum value at approximately 200 epochs and then increased, reflecting further overfitting. In other words, the re-creation of training data suppresses overlearning, contributing to the realisation of a more generalised model.

*2.3 Similar waveform search based on binary hash codes*

By applying the deep hashing network (Table 1) to a waveform, we can obtain an embedding vector and a 64-bit hash code through rounding. Furthermore, we can group waveforms with the same hash code by creating a hash table, wherein the waveform ID can be accessed from the hash code (i.e. the hash code is used as a key of the table). In the hash table, waveforms with a specific hash code can be searched without searching for other keys with a computational cost of $O(1)$, which is ideally independent of the size of the table. Furthermore, in previous studies, efficient methods have been proposed to perform nearest-neighbour searches even using a large hash table or hash code with larger dimensions (Liu et al. 2011; Norouzi et al. 2014).

We can also control the similarity of the grouped waveforms by grouping them based on a distance metric (the Hamming distance is typically used) between the hash codes rather than selecting the same code. Although this analysis requires distance calculations for all combinations of the analysed hash codes, a much lower computational cost is incurred compared with the calculations of cross-correlation coefficients among the original waveforms owing to the highly reduced dimensions. In the present case, we hashed a 1024-sample waveform, which was handled as single-precision (32-bit) real numbers after the multiplication of the gain factor, to a 64-bit binary code, resulting in a 1/512 compression of the data size.

## 3. Catalogued event grouping with hash codes

### 3.1 Similar waveform grouping

We applied the trained deep-hash model to 16 ch M304A waveforms of the 6057 events listed in the KJG1810 AE catalogue developed by Tanaka et al. (2021) and obtained their embedding vectors. Fig. 3 presents the values constituting the obtained vectors. As indicated, most of these values were close to zero or one, indicating that the quantisation loss (eq. (2)) functioned well. We obtained 64-bit hash codes by rounding these vectors and then constructed hash tables for similar waveform grouping.

We constructed 16 hash tables corresponding to each channel waveform. As an example, we present the results of the Ch 7 records, for which 5981 hash values were obtained. Of the 5981 values, 67 had two or more (four is the maximum) members with the same hash codes. Fig. 4 presents an example of these groups with the Ch 7 waveforms. In each panel, the average of the cross-correlation coefficients $Cc$ among all possible pairs in each group are indicated. Fig. 4(a) presents the maximum group with an average $Cc$ value of 0.97. Figs. 4(b)–4(e) present the groups with the maximum average $Cc$ value among the groups with Hamming distances of $d_h$ = 2, 11, 20, and 28 from the group illustrated in Fig. 4(a). Evidently, similar waveforms were successfully grouped in these cases. As $d_h$ increased, waveforms that appeared to be different from those shown in Fig. 4(a) could be extracted.

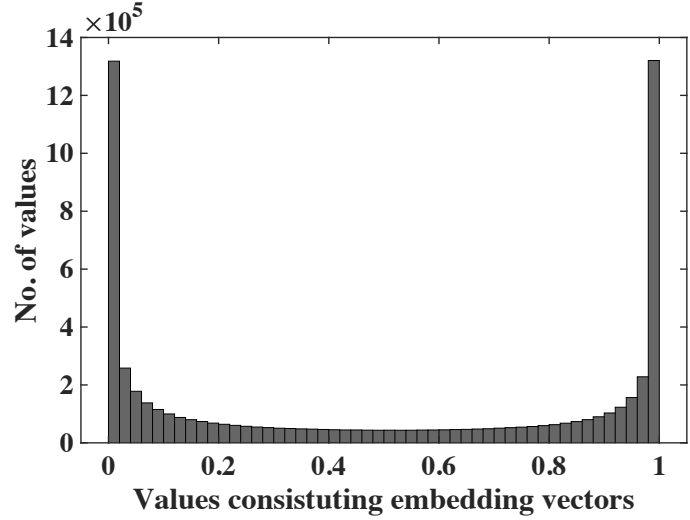

Fig. 3. Histogram of components constituting embedding vectors obtained by the deep-hash network. The embedding vectors are obtained for the 16 ch M304A records of the 6057 AE events listed in the KJG1810 catalogue developed by Tanaka et al. (2021).

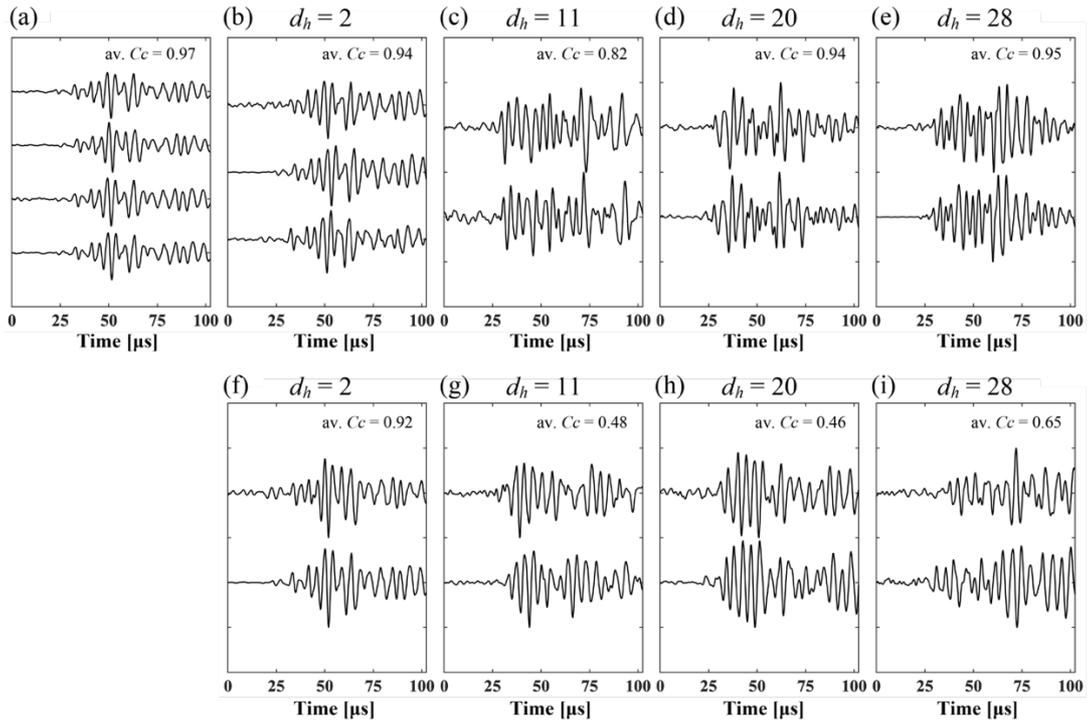

Fig. 4. Ch 7 waveforms of the events belonging to groups with the same hash code. (a) Results for the maximum group, in which the average $Cc$ value among all Ch 7 waveforms in a group is 0.97. (b)–(e) Results for groups with the highest average $Cc$ for $d_h$ = 2, 11, 20, and 28 from group (a). (f)–(i) Groups with the lowest average $Cc$ for each $d_h$.

Figs. 4(f)–(i) present the results with the lowest average $Cc$ in the group with the same $d_h$, as in (b)–(e). As illustrated in Figs. 4(g) and (h), waveforms with low $Cc$ values (below 0.5) could sometimes be grouped. Even for these low *Cc* pairs, similar features (e.g. phases with large amplitudes appearing at similar timings), likely caused by similar Green's functions, were identified, possibly increasing the number of detected events in similar waveform search problems. However, the waveforms of distant hypocentre pairs (e.g. several tens of millimetres apart) were often selected when those with the same hash codes were grouped (Fig. 5a), indicating erroneously grouped event pairs that were not theoretically predicted to have similar waveforms.

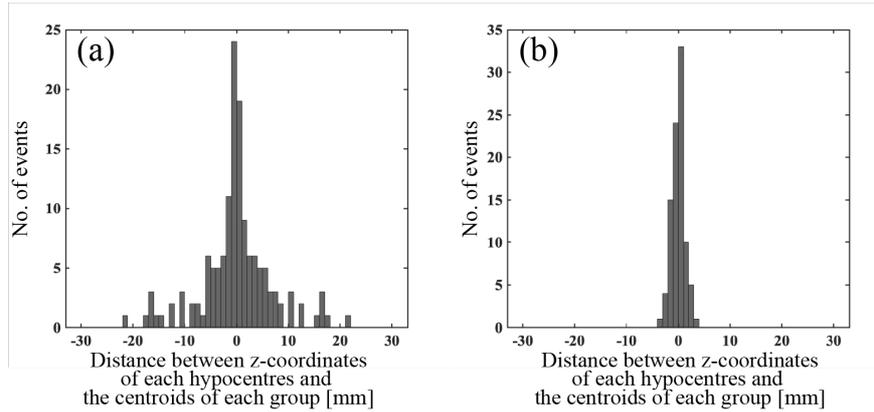

Fig. 5. Histogram of differences in the *z*-coordinates between the hypocentres and the centre of gravity of each group with the same hash codes. (a) Result for groups with the same hash codes for the Ch 7 records (standard deviation was 6.8 mm). (b) Grouping results for pairs with $d_h^{av}$ lower than 4.0 (standard deviation was 1.2 mm).

### *3.2 Relation between hash codes and cross-correlation coefficients*

As the cross-correlation coefficient *Cc* is the most popular indicator of waveform similarity in seismology, examining the grouping results and $d_h$ among the hash codes based on *Cc* is important. Fig. 6(a) illustrates the relationship between $d_h$ and *Cc* for all possible pairs of the Ch 7 waveforms of the 6057 events in the KJG1810 catalogue. The results present significant correlations (correlation coefficient of −0.1120 with a 95% confidence interval is [−0.1124, −0.1115]), and the average *Cc* drops rapidly as $d_h$ increases. As depicted in Fig. 7(a), the histogram of *Cc* for $0 \leq d_h < 4$ has a significant peak in *Cc* > 0.9. This peak decreases in intensity, and the fraction of pairs with high *Cc* values decreases rapidly as $d_h$ increases (Fig. 7b-e). As described in Section 2.2, we created positive pairs not only from event waveforms but also from noise waveforms. Figs. 7(f)– 7(j) present the same plots as Figs. 7(a)–(e) based on the model trained without noise waves, where the number of false-positive pairs (pairs with small $d_h$ and low *Cc*) increased, particularly in the small $d_h$ range. The use of triples between noise waves, which present low *Cc* values despite the similarity of the overall waveform characteristics, likely reduces the number of false positives.

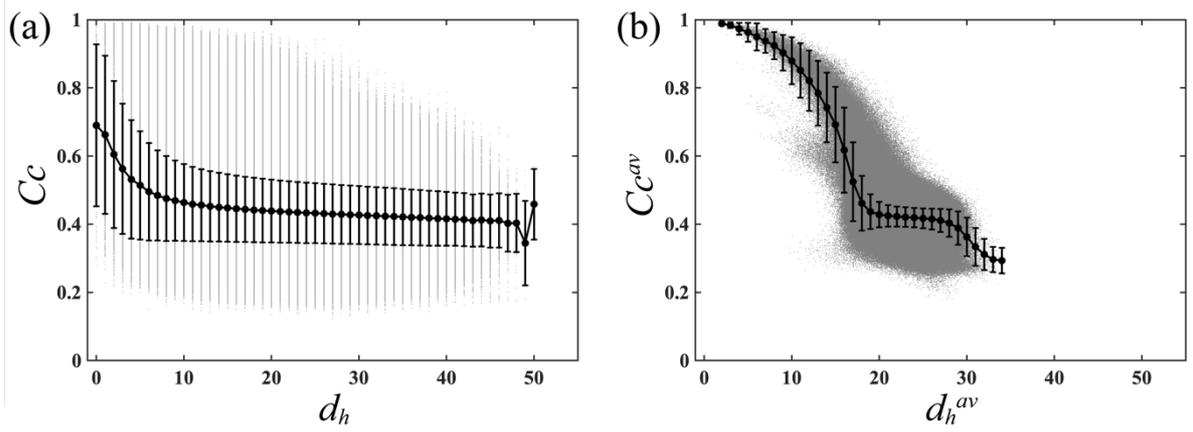

Fig. 6(a). Relationship between $d_h$ and $Cc$ for the Ch 7 waveform of all event pairs in the KJG1810 catalogue. Grey dots indicate the cross-correlation coefficients obtained for each event pair. Black dots indicate the average $Cc$ at each $d_h$, and the black lines above and below the average indicate standard deviation. (b) Relationship between $d_h^{av}$ and $Cc^{av}$, which are the 16 ch averages of $d_h$ and $Cc$, respectively.

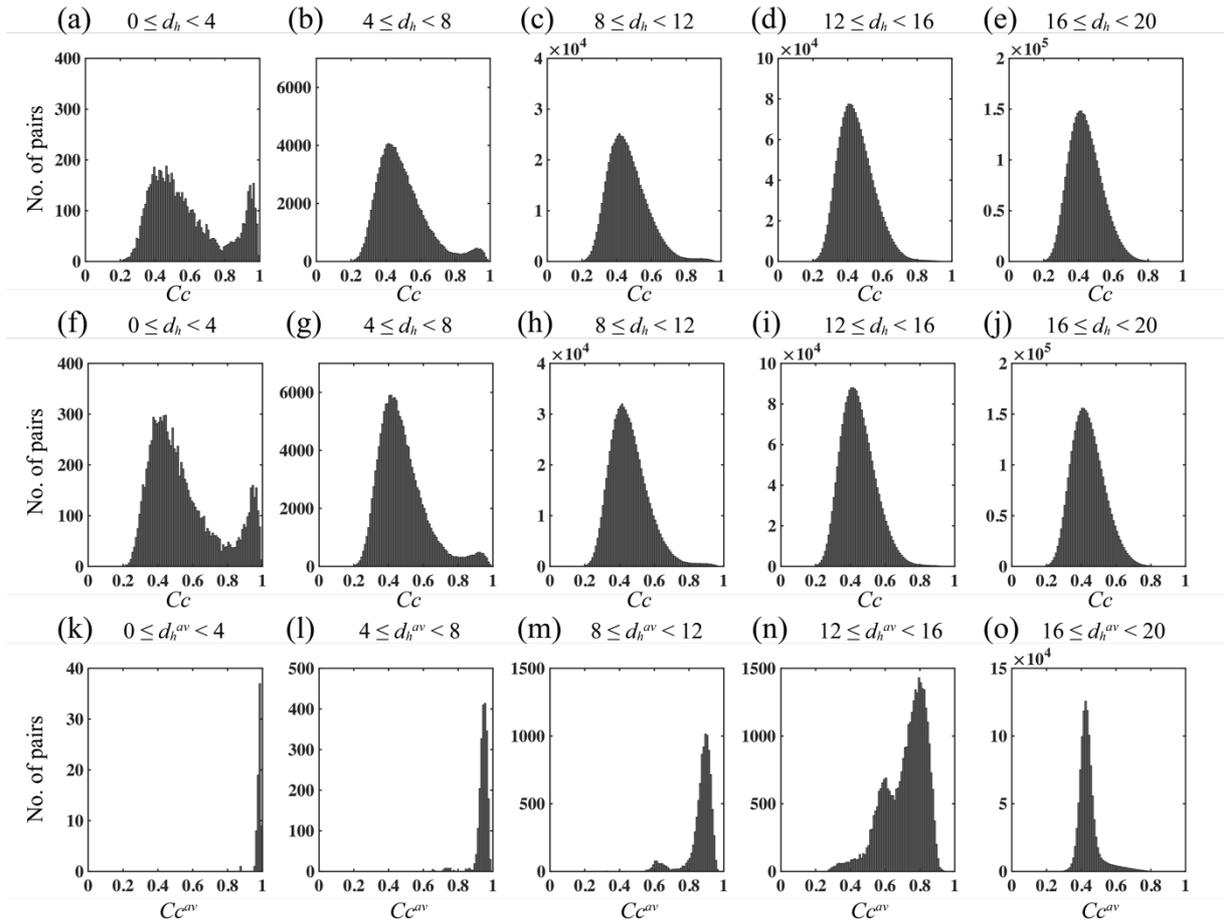

Fig. 7(a)–(j). Histograms of $Cc$ values evaluated for individual event pairs in the KJG1810 catalogue. Results for channel 7 records in specific ranges of $d_h$ are shown. (a)–(e) Results for the Hash function trained by triplets with event and noise waveforms. (f)–(j) Same plots for the case without noise triplets. (k)–(o) Histograms of $Cc^{av}$ for event pairs with specific ranges of $d_h^{av}$.

### *3.3 AE event grouping using average Hamming distances*

Figs. 6(a) and 7(a)–(e) indicate that numerous false positives (a small $d_h$ with a low $Cc$ pair) and false negatives (a large $d_h$ with a high $Cc$) remained when the waveforms of individual channels with the same hash codes were grouped. Similar to the event grouping based on $Cc$, we could reduce the rates of false positives and negatives by selecting event pairs with similar hash codes at multiple stations. In this study, event pairs were grouped using the average Hamming distance $d_h^{av}$ from the 16 M304A records. Fig. 6(b) presents the relation between $d_h^{av}$ and $Cc^{av}$, the average maximum $Cc$ among the 16 channels. The time lags exhibiting the maximum $Cc$ values were estimated separately for individual channels. The scattering considerably decreased owing to averaging, and the histograms of $Cc^{av}$ in a specific range of $d_h^{av}$ presented sharper peaks (Figs. 7k–o), corresponding to a reduction in the number of false positives and negatives. For the 6057 events in the KJG1810 catalogue, we attempted to group event pairs satisfying $d_h^{av} \leq 4.0$, obtaining 30 similar event groups consisting of 2–17 events. A total of 93 events belonged to one of the groups. This treatment also suppressed scattering in the hypocentres within each group (Fig. 5b), indicating a reduction in misdetections. Fig. 8 presents an example of the waveforms of the 17 events belonging to the largest group. Events with similar waveforms across multiple channels were successfully extracted.

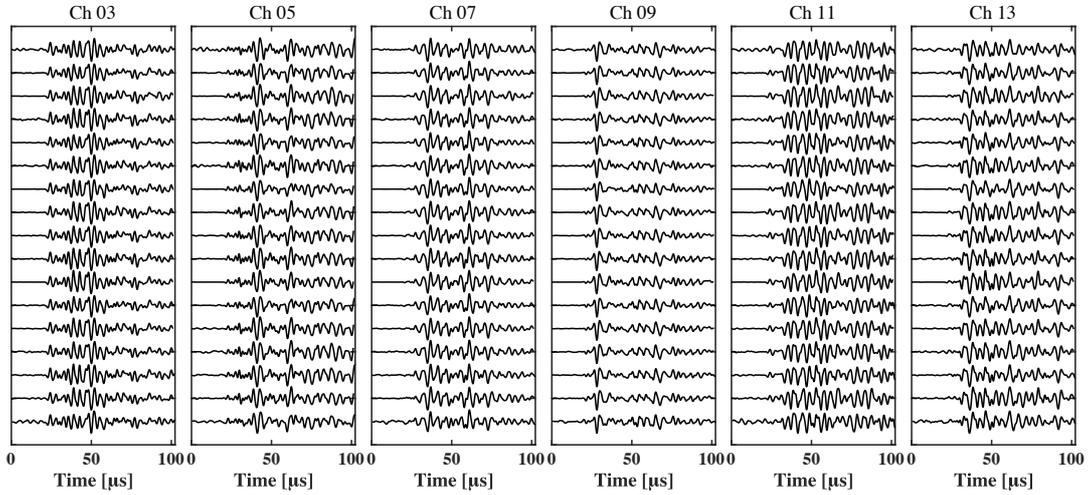

Fig. 8. Example of waveforms (Ch 3, 5, 7, 9, 11, and 13) for grouped events satisfying $d_h^{av} \leq 4.0$.

## 4 Application to the template matching problem

In this section, we address the template matching problem for the records of the KJG1810 experiment using the deep hashing model trained previously (Section 2). We applied the model to the continuous waveform records to obtain their hash codes, following which we extracted windows with small $d_h^{av}$ values against the 6057 template events listed in the KJG1810 catalogue. For this analysis, we separated a 10 MHz sampling continuous waveform at each channel into 35,128,124 windows, which were 1024 samples in length with a 50% overlap with the neighbouring windows. The generation of Hash codes for all the windows took 3.5 hours when using an NVIDIA A40 GPU. For the KJG1810 experiment, approximately 30 min waveforms were recorded, and the number of samples reached $1.8 \times 10^{10}$ at each channel, corresponding to the 5.8-year record for a 100 Hz sampling rate typically adopted in surface seismic observations.

Fig. 9(a) presents a histogram of $d_h^{av}$ between the 6057 template events and 35,128,124 windows. The histogram exhibits a bimodal distribution with peaks around $d_h^{av} \sim 2$ and $d_h^{av} \sim 30$. Of the windows exhibiting $d_h^{av} < 4$ in minimum against the templates, 99.6% correspond to redetection of the template events. The grouping analysis in Section 3.3 (the criterion of $d_h^{av} < 4$ was adopted) was performed for waveforms comparable to the refinding of the templates.

We detected events in the continuous records by searching for windows with $d_h^{av}$ lower than a threshold. The strict criterion of using a smaller threshold reduced both the number of detected events and false positives, similar to the template matching analysis based on $Cc$. For example, Kato et al. (2012) extracted the results of $Cc > \mu + 8\sigma$, where $\mu$ denotes the average and $\sigma$ denotes the standard deviation of $Cc$ in the template matching analysis. In this study, we extracted the event candidate windows using a looser criterion of $\mu - 6\sigma$ based on the average and standard deviation of $d_h^{av}$, which corresponds to the selection of $d_h^{av} < 17.48$ (please note that the sign of the second term is changed because $Cc$-based analysis extracts windows with high $Cc$ values, whereas $d_h^{av}$-based analysis extracts small $d_h^{av}$). After extraction, false positives were removed using the lag times between the templates and detected events, which were estimated using a cross-correlation technique.

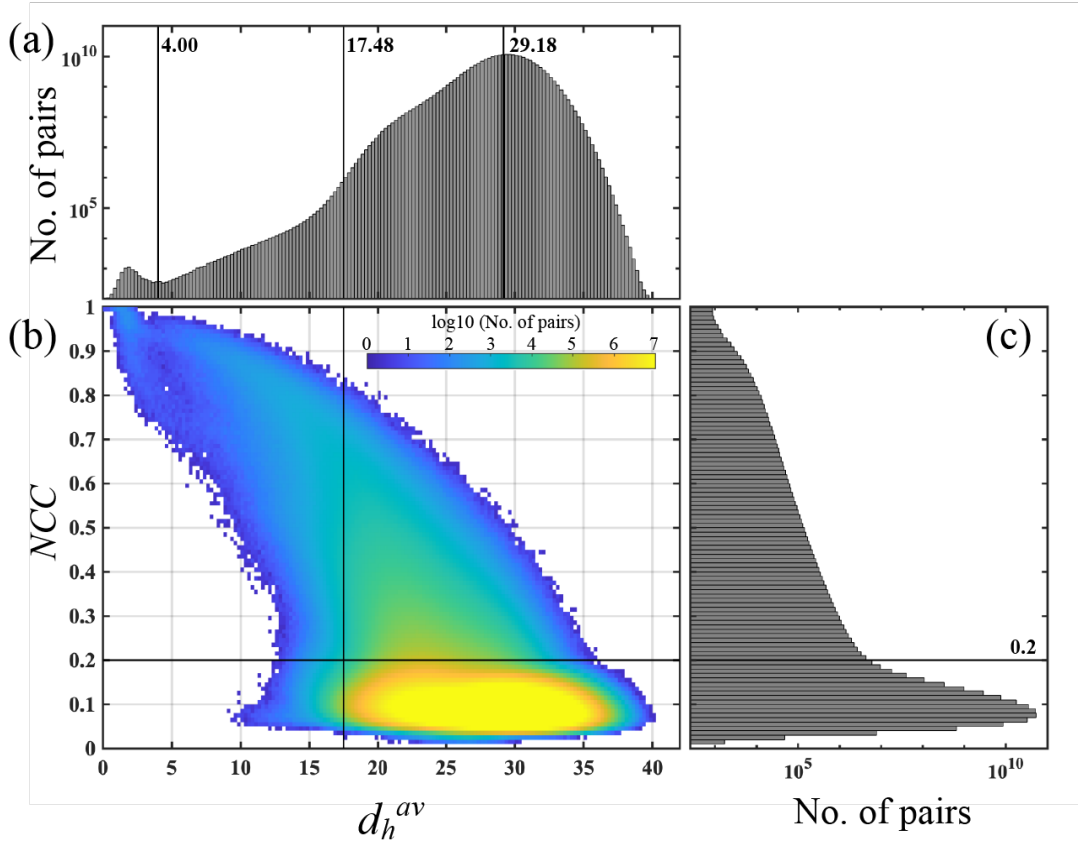

Fig. 9. (a) Histogram of $d_h^{av}$ between the 6057 templates and all 35,128,124 windows for the continuous waveform records of the KJG1810 experiment. The average and standard deviation of the obtained $d_h^{av}$ were 29.18 and 1.95, respectively. (b) Relationship between $d_h^{av}$ and $NCC$. (c) Histogram of $NCC$ between the same template-window pairs. The average and standard deviation of the obtained $NCC$ were 0.083 and 0.016, respectively.

Of the 35,128,124 windows, 424,457 (~1.21%) have one or more templates with $d_h^{av}$ < 17.48. We calculated the lag time $dt$ resulting in the maximum $Cc$ between the waveforms in the event candidate windows and the corresponding template waveforms (i.e. the waveforms of template events showing the minimum $d_h^{av}$). When an event near the template event was detected appropriately, the scattering in $dt$ among all channels approached zero. Fig. 10(a) illustrates the relationship between $d_h^{av}$ and $dt^{MAD}$, the median absolute deviation of $dt$ among the 16 ch records, for the 424,457 pairs. Evidently, $dt^{MAD}$~0 was obtained for a small $d_h^{av}$, whereas it increased with increasing $d_h^{av}$ and approached large values particularly for $d_h^{av}$ > ~10. Here, we selected 46,502 windows satisfying $dt^{MAD}$ < 1.5 μs as the windows containing an event with consistent travel time differences from a template event. Incidentally, although we conducted a high-cost calculation of $Cc$, the computation cost was reduced to 0.0002% compared to when the $Cc$ values were calculated among all templates and windows, which is necessary for usual template matching analysis based on $Cc$.

For the 46,502 selected windows, we estimated the origin times of the detected events from $dt^{MAD}$ and the origin time of the template event. We regarded origin times within a 50 μs difference as duplicated detection and merged them into one event, eliminating 16,983 detections. The remaining 29,519 windows included all 6057 template events, and the remaining 23,462 could be regarded as newly detected events in this analysis.

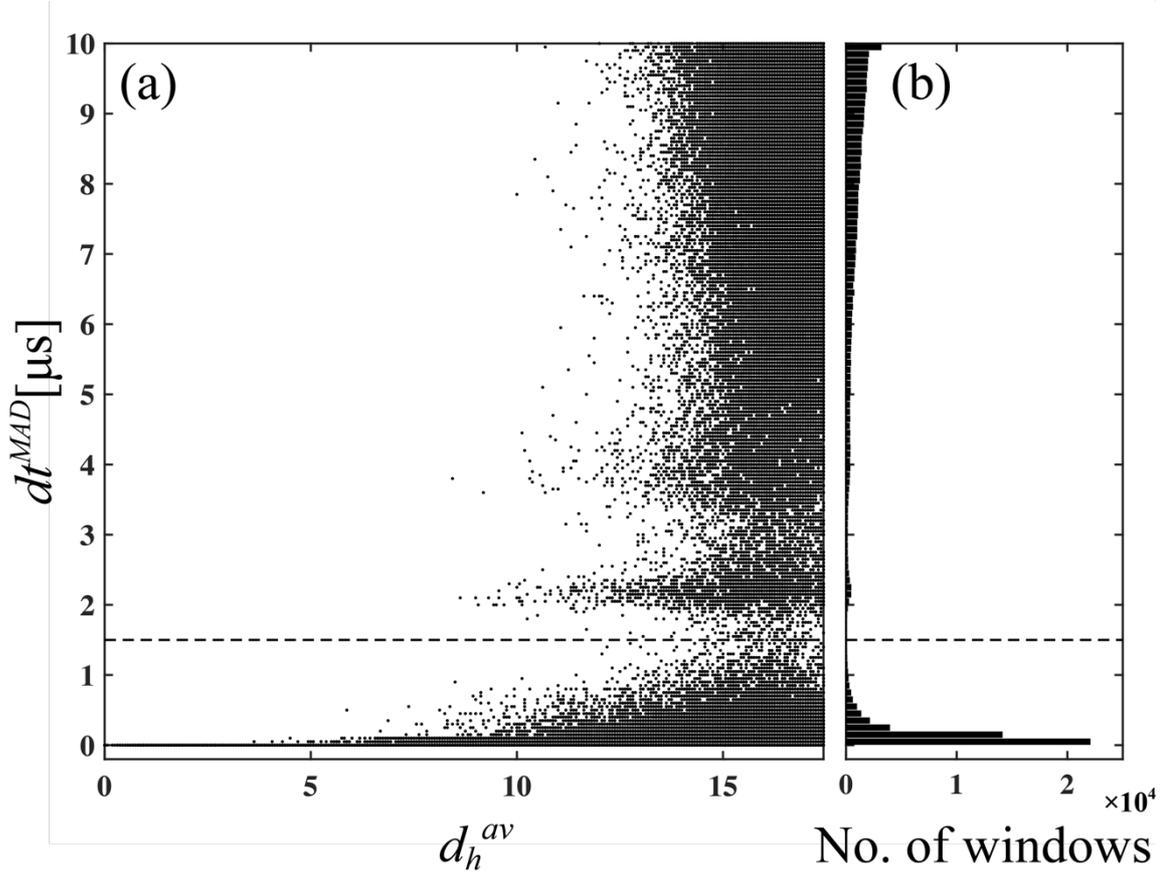

Fig. 10(a). Relation between $d_h^{av}$ and $dt^{MAD}$ for the 424,457 pairs between selected windows and corresponding templates in the hashing-based template matching. (b) Histogram of $dt^{MAD}$.

Fig. 11 presents examples of the waveforms of newly detected events in hashing-based template matching for the KJG1810 experiment. Notably, waveforms exhibiting high correlations with the templates were detected even for windows with low signal-to-noise ratios, which resulted in indistinguishable P wave onsets. Even for large $d_h^{av}$ such as > 17, the event detection process appeared to function appropriately (Fig. 11d). Fig. 12 presents a continuous waveform when the event occurrence rate is high immediately before the breakdown (a rapid drop in fluid pressure due to macroscopic fracture generation). The occurrence times of the template events and events detected by the hash-based method are also indicated. As shown in this figure, hashing-based template matching detected numerous events at the time of transient amplitude changes.

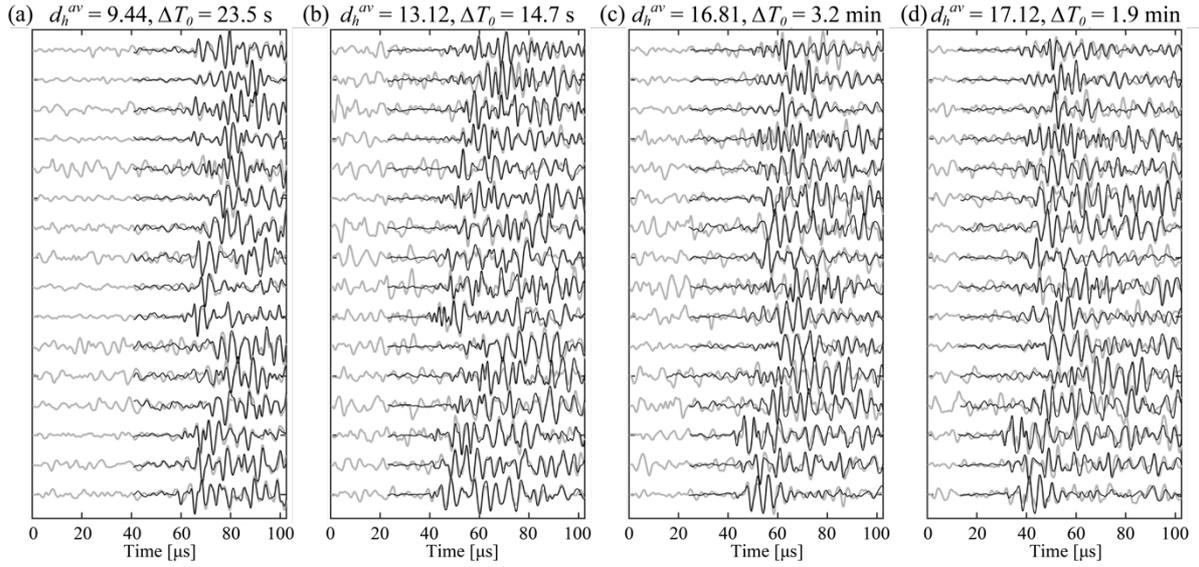

Fig. 11. Example of newly detected events in hashing-based template matching for continuous records of the KJG1810 experiment. Black lines represent the template waveforms, and grey lines represent the newly detected ones. We plot the template waveforms of each channel after moving to the timings exhibiting maximum waveform correlations against the detected waveforms. $\Delta T_0$ in the figure indicates the time difference between the template waveform and the detected waveform.

We located the hypocentres of the obtained 29,519 events based on the double-difference method (Waldhauser & Ellsworth 2000) using the cross-correlation travel time reading technique. For the 23,462 newly detected events, we assumed the same initial hypocentres as those of the corresponding template events. For the maximum of the 300 neighbouring events within 20 mm from the initial hypocentres, we estimated lag times maximising $Cc$ for $P$ and $S$ waves using 0.2–1.2 MHz bandpass filtered waveforms with a length of 25.6 μs around theoretical arrivals. The travel time differences for the pairs exhibiting $Cc \geq 0.65$ were used for the relocation. For events catalogued by Tanaka et al. (2021), we also used the travel time differences calculated from the $P$ arrival-time readings used in their cataloguing procedure. We successfully obtained 25,692 hypocentres (87.0%) using this localisation procedure.

Fig. 13 presents the $z$-coordinates of the hypocentres obtained against the time from the breakdown. Tanaka et al. (2021) pointed out that two expansion phases of AE activities can be recognised before the breakdown in an experiment with Kurokami-jima granite. In the KJG1810 experiment, the first phase started from $t \sim -400$. The active region gradually expanded outward from the borehole at $z = 0$. The second phase began at $t \sim -40$ s, and their active region with higher density expanded more rapidly from the pressurised section along the $z$-axis. Tanaka et al. (2021) interpreted that the first phase was induced by fluid penetration through pores or preexisting cracks, and the second phase corresponded to the main fracture formation. The increased number of AE hypocentres with hashing-based template matching revealed these phases more clearly, particularly in the second phase.

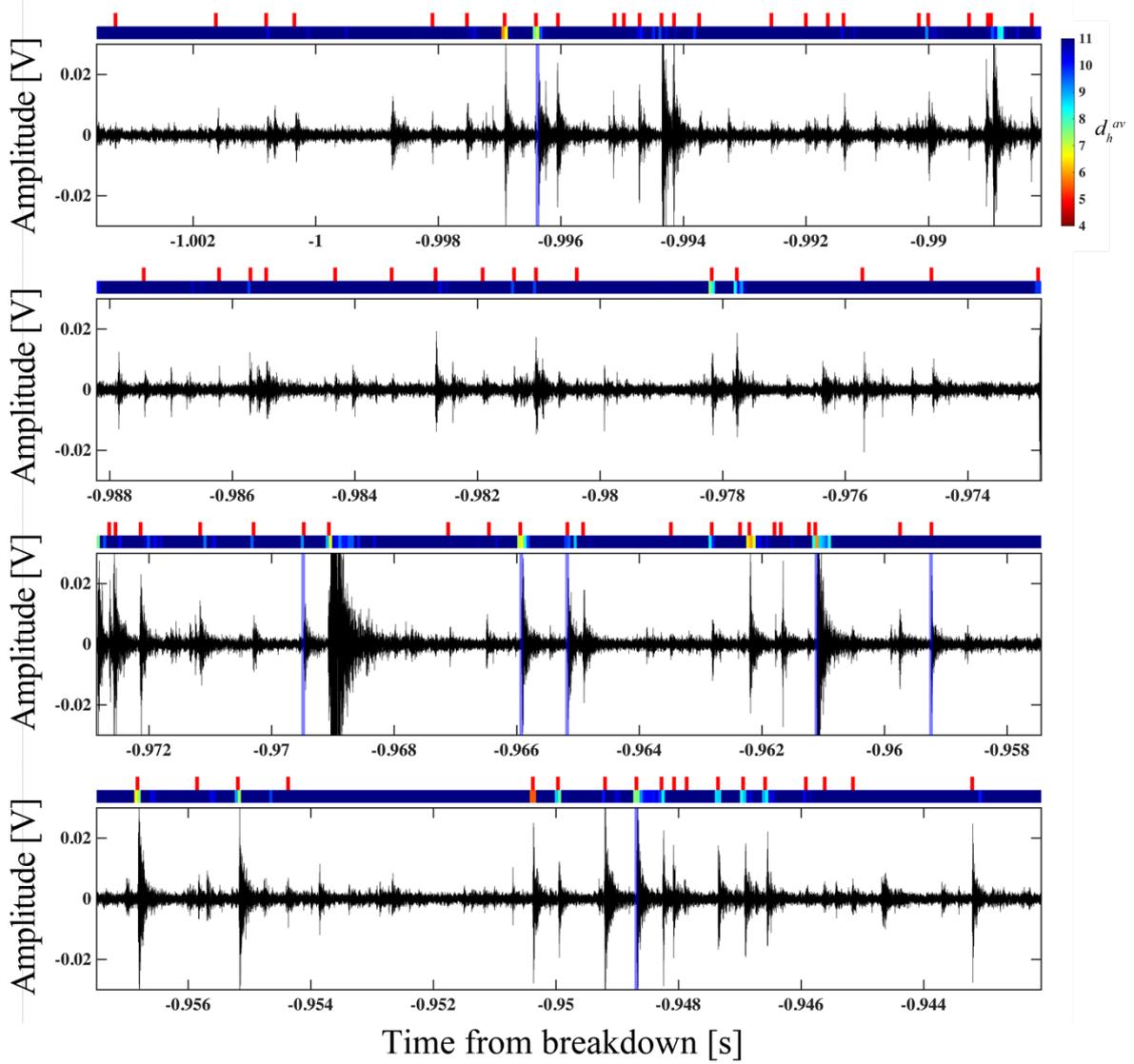

Fig. 12. Continuous waveform of Ch 1 recorded ~1 s before the breakdown in the KJG1810 experiment. The vertical blue lines depict the event occurrence times of the template events catalogued by Tanaka et al. (2021). The red lines above the waveform indicate the occurrence times of the events detected in hashing-based template matching. Minimum $d_h^{av}$ obtained by the autocorrelation analysis in Section 5 is presented using a colormap above the waveform.

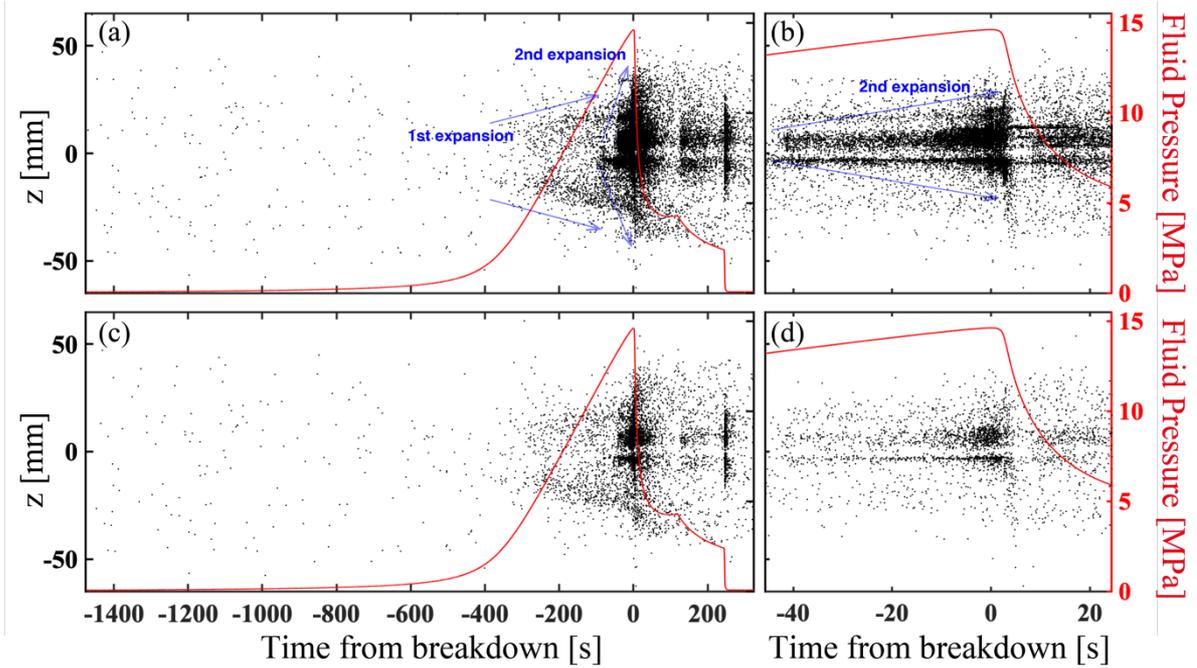

Fig. 13. Temporal change in the *z*-coordinates of the AE hypocentres obtained for the KJG1810 experiment. The breakdown time (the origin of the horizontal axis) is defined as the timing of the peak pressure. (a), (b) Relocated results by the double-difference algorithm for the 25,692 events obtained by hashing-based template matching (Section 4). (c), (d) Relocation results for 6055 events included in the original catalogue developed by Tanaka et al. (2021). (b) and (d) show the enlarged views around the breakdown ($t = 0$) of (a) and (c), respectively.

We also solved the template matching problem using a standard approach based on *Cc* to compare the detection performance. We calculated the network cross-correlation coefficient (*NCC*; Gibbons & Ringdal 2006) for the same event-window pairs (i.e., all combinations of the 6057 templates and the 35,128,124 windows). Fig. 9(b) presents the relation between $d_h^{av}$ and *NCC*, and Fig. 9(c) presents a histogram of *NCC* for all pairs. Fig. 9(b) shows a strong peak in the area of $d_h^{av}$ = 20–35 and *NCC* = 0–0.15, likely indicating unrelated waveform pairs. The selection of $d_h^{av}$ < 17.48 in the hashing-based analysis successfully excludes this peak. However, pairs with high *NCC*~0.8 in the maximum are involved in the excluded pairs. This indicates that our hashing method does not have enough capability to separate high and low *NCC* pairs, resulting in many false detections, even after averaging 16-channel $d_h$. This highlights the necessity to improve the hash function performance (we will further discuss this problem in Section 6.3.1). We can obtain a larger number of well-relocated hypocenters based on the *NCC* approach. As for the *NCC*, we selected 755,606 event candidate windows accompanying one or more templates exhibiting *NCC* > 0.2. This threshold corresponds to $\mu_{NCC}$ + 7.25$\sigma_{NCC}$, where $\mu_{NCC}$ is the average and $\sigma_{NCC}$ is the standard deviation of the maximum *NCC* for each window pair. The obtained histogram of NCC presents a significant kink at this value (Fig. 9c), which likely represents the deviation from the distribution of *NCC* between unrelated waveforms (Aso et al. 2011 and Hirano et al. 2022). For the waveforms in the candidate windows, we estimated origin times based on time lags evaluated by *NCC* calculations. We then removed duplicated detections, read arrival time differences, and relocated their hypocenters using the same procedure of the hashing-based analysis. We finally obtained 63,702 events, which indicated 2.5 times as many hypocenters obtained by the hashing-based approach.

Although our deep hash method is still required for performance improvement, its low computational cost based on distance calculation between compact binary codes is attractive. We compared the intrinsic computing time between $d_h^{av}$ and *NCC* calculations (i.e., swap-in/swap-out time, which is extensive for *NCC* calculations, is not considered because it highly depends on hardware settings). According to our benchmark test under 120 thread parallelisation using an AMD Ryzen Threadripper 3990X, the *NCC* calculation of the same window pairs required 1000 times higher computational costs under the implementation described in the Appendix, despite assuming a situation in which all waveforms were stored in the computer memory. Incidentally, when the original waveform data are handled as single-precision real numbers, their data size is 512 times greater than that of the corresponding hash codes, and the *Cc* calculation is more complicated than that of $d_h$, where only 'xor' and 'popcount' operations are required. These facts indicate that a drastic reduction in the computational cost of *Cc* is difficult, although the computing cost of *Cc* calculations relies on the implementation.

The time required to read waveform data into a computer memory further emphasizes the advantages of the hashing-based approach, especially when a large dataset is analysed. For example, the *NCC*-based template matching using a template dataset larger than computer memory requires a repetitive swap-in/swap-out for the templates. The extreme case is the autocorrelation problem described in Section 5, and it often requires unrealistic computing time. The deep hashing approach can avoid this problem because of the small resultant binary codes.

## 5 Application to the autocorrelation problem

In Section 4, we calculated the 64-bit hash codes of all 35,128,124 windows for each M304A channel to conduct hashing-based template matching for continuous records of the KJG1810 experiment. The total size of the obtained hash codes was 4.5 GB (64 bit x 35,128,124 windows × 16 channels), which can be stored in computer memory in a typical, recent computing environment. The highly reduced data size from that of the original waveforms allowed us to calculate $d_h$ for all combinations among those windows, corresponding to a hashing-based autocorrelation analysis. Under the same computing environment of *NCC* calculations in Section 4, $d_h$ calculations for all combinations was accomplished only in 15.5 h. We identified a window exhibiting the minimum $d_h^{av}$ value against each reference window (hereinafter referred to as the 'minimum-$d_h^{av}$ window'). In this identification, neighbouring windows overlapping with the reference window were excluded.

Fig. 14 presents a histogram of the obtained $d_h^{av}$. The distribution is bell-shaped, with an average of approximately 20.7, likely reflecting background noise correlations. In the double logarithmic plot (Fig. 14a, inset), a significant kink is observed at $d_h^{av}$ ~10. This change in the $d_h^{av}$ distribution shape is likely caused by the occurrence of similar events, and it is also observed in the autocorrelation analysis based on *Cc* (Aso et al. 2011; Hirano et al. 2022).

Fig. 15(a) presents the minimum $d_h^{av}$ window ID and the corresponding $d_h^{av}$ values for all reference windows. Notably, small $d_h^{av}$ values were frequently obtained around the breakdown (window ID of ~2.88 x $10^7$), where the AE occurrence rate was very high. At the time of the breakdown, we identified a 'white band', indicating a lack of small $d_h^{av}$ values during the period. During and immediately after the breakdown, elastic waveforms accompanying the generation/propagation of a macroscopic fracture and numerous AE events were continuously recorded (Fig. 16), and the white band likely reflected the different characteristics of the waveforms between this and other periods. In Fig. 15(b), an enlarged view of the final part of the experiment, including the breakdown time, reveals that a smaller $d_h^{av}$ tends to occur between a pair of windows from before the breakdown, or between a pair of windows from after the breakdown. Such small $d_h^{av}$ values were rarely obtained for a pair of windows where one was before and the other was after the breakdown. This may indicate that the magnitude and focal mechanisms of AE events changed significantly before and after the breakdown (Tanaka et al. 2021; Naoi et al. 2022). In addition, small $d_h^{av}$ values tended to be concentrated along the diagonal line, indicating that similar events frequently occurred in temporally close windows. Note that the sudden drop in fluid pressure around the $3.36 \times 10^7$th window corresponds to an artificial release of fluid pressure. At this time, the reference windows with small $d_h^{av}$ values against the post-breakdown windows are concentrated.

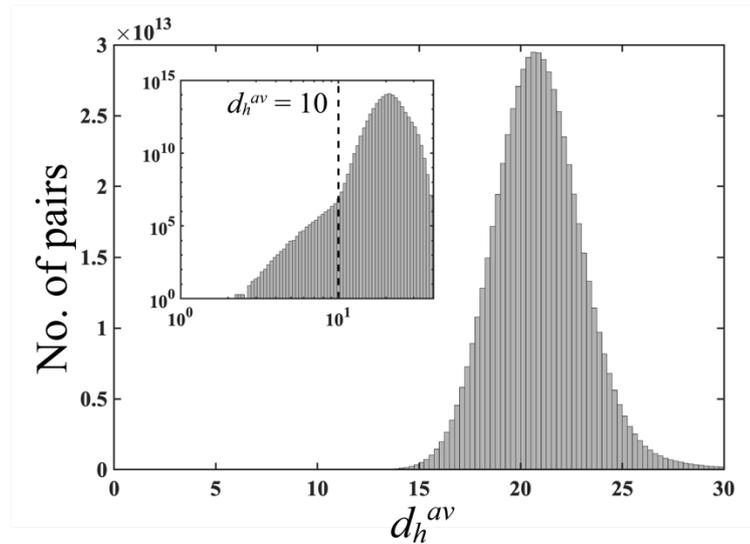

Fig. 14. Histogram of $d_h^{av}$ for all possible pairs among the 35,128,124 windows for the continuous waveform record of the KJG1810 experiment. (a) Histogram of $d_h^{av}$. (inset) Double-log plot of the same histogram.

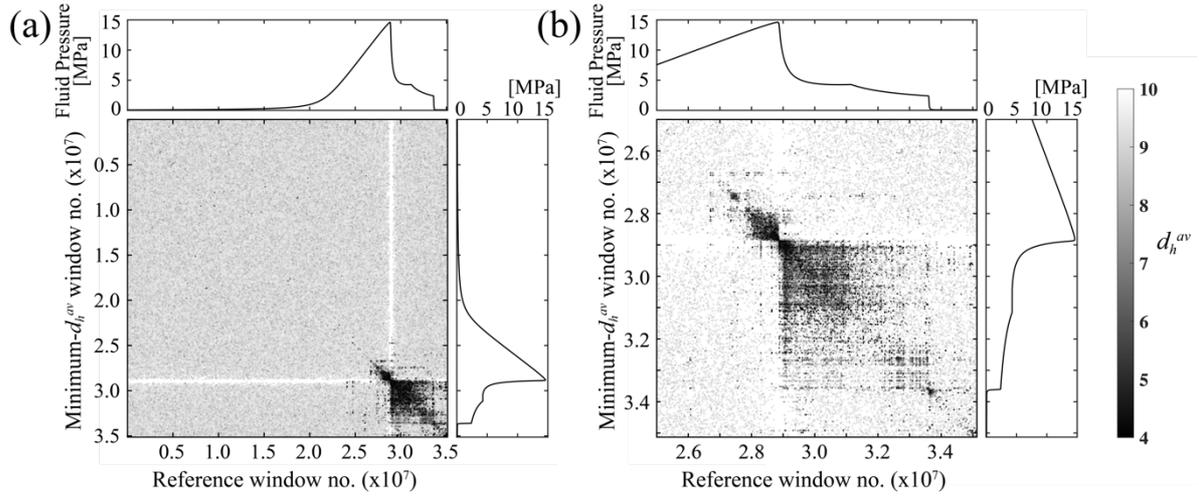

Fig. 15. Minimum-$d_h^{av}$ window ID and corresponding $d_h^{av}$ values for all 35,128,124 reference windows of the KJG1810 experiment. Adjacent windows overlapping with each reference window were excluded from the search of the minimum-$d_h^{av}$ windows. Owing to parallelisation, one of the minimum-$d_h^{av}$ windows was randomly chosen if reference windows had two or more windows with the same minimum $d_h^{av}$ values. We performed this calculation twice, and only 1.8% of all presented different results, indicating that this ambiguity does not affect the discussion in the text. (a) Results for all windows. (b) Enlarged view of the final part of the experiment.

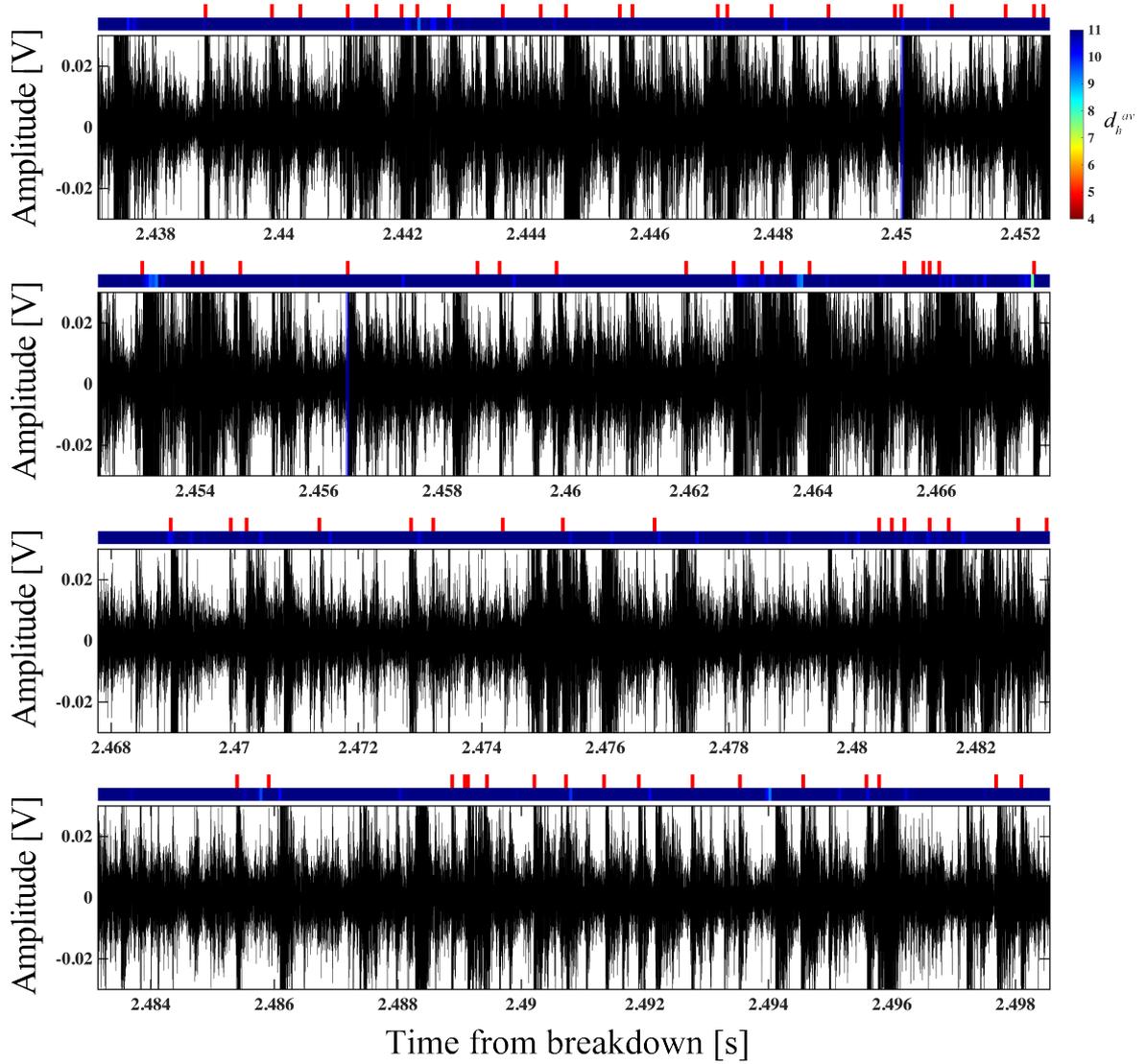

Figure 16. Continuous waveform of Ch 1 recorded ~2.5 s after the breakdown in the KJG1810 experiment. This period corresponds to the 'white band' in Fig. 15. The vertical blue lines depict the event occurrence times of the template events catalogued by Tanaka et al. (2021). The red lines above the waveform indicate the occurrence times of the events detected in hashing-based template matching. The minimum $d_h^{av}$ obtained by the autocorrelation analysis in Section 5 is presented using a colormap above the waveform.

In the *Cc*-based autocorrelation analysis, repeating signals that do not correspond to event waveforms, such as coda waves of large events or random/repeated noise, are often detected (Gibbons & Ringdal 2006; Rong et al. 2018). The same problem occurs in hashing-based analysis. Fig. 17 presents examples of windows satisfying $d_h^{av} < 10$. Although numerous windows containing transient amplitude changes likely corresponding to AE events associated with the rupture of rock were selected using this criterion (Fig. 17a), the coda waveforms of these AE events also presented a small $d_h^{av}$ (Fig. 17b). We frequently found pairs that were likely to detect highly correlated noise pairs (Fig. 17c). The selected windows often contained monotonic waveforms (Fig. 17d), which may correspond to the sound of a fluid flowing through pipes or cracks.

As demonstrated in Section 4, numerous waveform pairs with $d_h^{av} > 10$ correspond to similar event pairs, and they are missed if the detection criterion is determined from the kink of $d_h^{av} = 10$ (Fig. 14 inset). The minimum $d_h^{av}$ value obtained in the hashing-based autocorrelation for each window is depicted above the waveform plots in Figs. 12 and 16. Although Fig. 16 rarely represents small $d_h^{av}$ windows, probably due to specific characteristics in the white band period, the timings of many $d_h^{av} < 10$ windows in Fig. 12 were consistent with the events detected by the hashing-based template matching (red lines), whereas larger minimum $d_h^{av}$ values were also frequently obtained at the positions of red lines. When a greater threshold is adopted, additional quality control is necessary to extract AE events from candidates with $d_h^{av}$ values smaller than the threshold to eliminate numerous misdetections. This problem is essential to apply the deep hashing approach to the autocorrelation problem because a high computing cost in false positive removal might negate the advantage of the hashing-based method in computing cost. That is, our demonstration with a low computing cost showed only prerequisites to encourage further studies. Improvement of the loss function and applying a machine learning classification similar to that used in $Cc$-based template matching (Herrmann et al. 2019) may be effective to suppress the false positive rate. This investigation is beyond the scope of this study because it is too large a problem to treat in the present study.

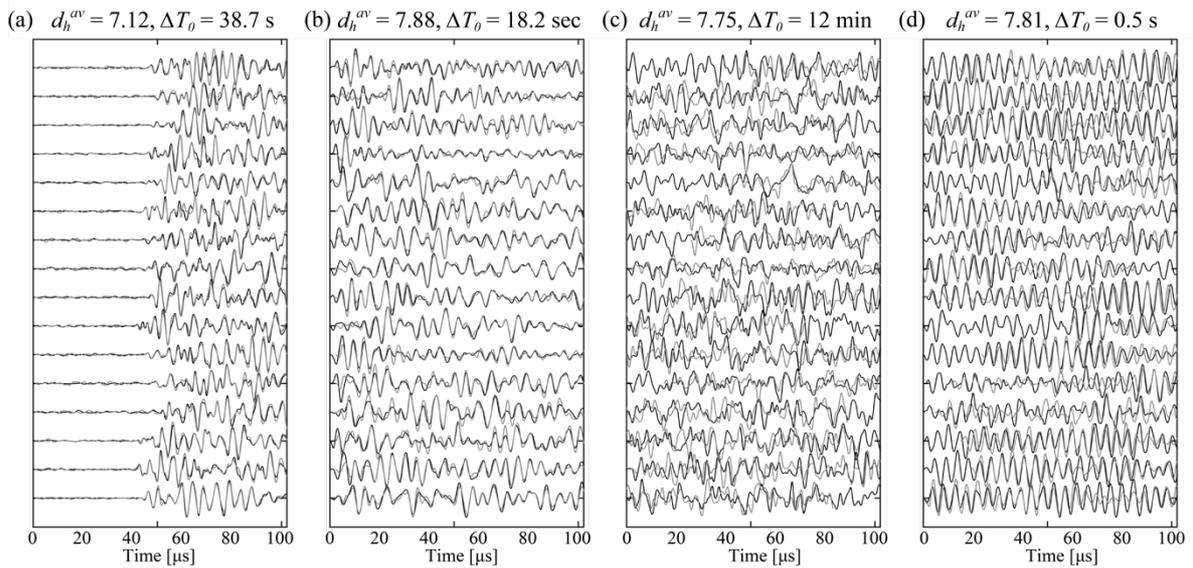

Fig. 17. Examples of similar waveforms identified by hashing-based autocorrelations for the KJG1810 experiment. The waveforms indicated by grey lines are plotted after time shifting to present the maximum $Cc$ against the waveform shown in black. The appropriate time shift was separately evaluated for each channel. Above each panel, $\Delta T_0$, the difference between the template window and the detected window, is indicated.

## 6. Discussions

### *6.1 Generalisation capability of the hash functions*

In Section 2, we described the training of a deep hashing model based on the AE catalogue of the KJG1810 experiment. In Sections 4 and 5, the model was applied to a continuous waveform record from the same experiment to detect additional events. In this case, the model possibly 'memorised' the catalogued waveforms using numerous deep learning model parameters. However, this could be problematic when solving the template matching problem using incomplete templates and autocorrelation problems.

We verified the applicability of the trained hash model for unknown waveforms, while the continuous decrease in validation loss for the KJG1811 dataset (Fig. 2a) indicated that this concern was not serious. We trained the deep hashing network based on a dataset obtained from an experiment using a different specimen, applied the resultant model to the KJG1810 record, and compared the detection performance in template matching analysis. Training was conducted based on the KJG1802 catalogue consisting of 8170 events (Tanaka et al. 2021) with the same number of noise waveforms recorded in the experiment. The same validation and test data were used in the training process as in Section 2.2 (i.e. based on the KJG1811 data). Using the resultant model, we obtained hash codes for waveforms of the 6057 events in the KJG1810 catalogue and the 35,128,124 windows cut from the continuous data, following which the $d_h^{av}$ values among them were calculated. The average and standard deviation of the obtained $d_h^{av}$ were 26.34 and 1.60, respectively.

Of the 35,128,124 windows, 869,066 satisfied the criterion of $d_h^{av} < \mu-6\sigma$ (16.74). We eliminated duplicate detections and misdetections according to the procedure described in Section 4, which resulted in the detection of 28,945 events (consisting of all 6057 templates and 22,888 new events). This comparable number of final detections with those detailed in Section 4 (29,519) indicates that the deep hashing model did not simply memorise the waveforms but also learned an effective mapping that extracted waveform features. Although these numbers are affected by randomness in the training process (initial weight parameters in the deep learning model and random selection in the triplet construction), we performed the same analysis five times for each training dataset and obtained comparable average numbers of the final detections (26,082 with a standard deviation of 4113 for the KJG1810 training data and 26,785 with a standard deviation of 1366 for KJG1802).

In the data-driven hash learning process conducted in this study, performance reduction owing to incomplete templates may be an issue (Bergen & Beroza 2019). For example, in the application of earthquake observation data, it is important whether a hash function trained using a dataset available at a specific time will be useful for future unknown earthquakes. Although this problem should be carefully verified in actual applications, the obtained hash functions operate well for unknown waveforms that were not used in training, at least in this study.

*6.2 Required memory of deep hashing approach*

As described in Section 1, Yoon et al. (2015) developed the FAST algorithm employing a simpler hash function for similar waveform searches. The advantages of our method over the FAST algorithm are smaller resultant hash codes and simpler postprocessing, enabling efficient use of computer memory.

In the FAST algorithm, Yoon et al. (2015) created a 4096-bit binary fingerprint from a waveform spectrogram and applied Min-hash (a function returning the smallest index containing one after a random permutation of the fingerprint) for a similar waveform search. Because this simple hashing possibly produces the same results for non-similar fingerprints, numerous hashings with different random permutations are applied to a fingerprint. A pair is judged to be similar when the matching rate exceeds a predetermined criterion to control accuracy. In a demonstrated example, they applied 700 hashings to a fingerprint and obtained the same number of 8-bit hash codes, resulting in a total of 5600-bit codes for a fingerprint. In this algorithm, hash tables were created and stored in the memory for all 700 hashes. Yoon et al. (2015) indicated that 36 GB of memory was required to analyse 1.6 x $10^7$ windows extracted from one channel waveform recorded at a sampling rate of 100 Hz during a six-month period. However, in the proposed deep-hash approach, only 0.13 GB of memory is necessary to save hash codes for the same number of windows, and minimal additional overhead (such as the preparation of many hash tables) is required. Actually, we performed a similar event search by placing all hash codes for many more windows ($5.62 \times 10^8$ windows = 35,128,124 windows × 16 channels) in memory. As for the FAST algorithm, efficient optimisation of fingerprint generation (Bergen & Beroza 2019), suppression of output size by result integration in multiple channels/stations (Bergen & Beroza 2018), use of Min–Max hash, partitioning of the hash database, and introduction of several domain-specific filters (Rong et al. 2018) enabled application to 11 stations (27 channels), 6–11 year continuous records of surface seismic observations, on a Linux server with two 28 thread CPU and 512 GB memory (Yoon et al. 2019). The proposed deep-hash method can be applied to large datasets using fewer computational resources.

*6.3 Remaining issues*

6.3.1. Improving the performance of the hash function

In event detection problems, the robustness of a hash function to time shifts (i.e. similar hash codes are obtained for different time shifts) reduces computational costs because it allows the realisation of moving windows with large time steps and facilitates the applications of long, continuous waveform records. Yoon et al. (2015) attempted to achieve the foregoing in their FAST algorithm by computing hash codes based on spectrograms and adopting moving windows with 90% overlaps with the neighbouring windows. In this study, we obtained a hash function with such robustness via training using positive pairs formulated based on waveforms extracted at randomly perturbated timings. Fig.

18 (a) presents the change in $d_h$ between a template waveform and another highly correlated waveform (the target waveform). We extracted the target waveform based on the movement of several windows at an interval of 2 μs and obtained similar $d_h$ for all windows, indicating the robustness of the hash function against timing differences. However, in the template matching problem, the detection performance improved slightly when smaller time steps were adopted. For example, when using 75% (256 samples) overlapping windows for the continuous waves, the total detected numbers obtained using the model trained on the KJG1810 dataset (Section 4) increased by ~10% (32,767 events). Improving the robustness of a hash function to time shifts contributes to more efficient calculations.

Another problem was the inconsistency between $d_h$ and $Cc$. As indicated in Fig. 6(a), the obtained hash function generated numerous false positives (a small $d_h$ for a low $Cc$ pair) and negatives (a large $d_h$ for a high $Cc$ pair). Although we mitigated the inadequate performance of the hash function by calculating $d_h^{av}$, the comparison between $NCC$ and $d_h^{av}$ in the template matching problem (Section 4) indicated the necessity of further improvement of the hash function.

Reduction of the false positive rate mainly contributes to the analysis efficiency. It reduces the calculation costs of quality control (e.g. the $dt^{MAD}$ analysis in Section 4) and the required dimensions of the binary code. However, even if false positives remained, we can use the proposed deep hashing method for pair screening to reduce the costs of subsequent $Cc$ calculations. That is, $Cc$ calculation is necessary only to remove false positives after selecting a pair with small $d_h$.

The existence of false negatives (a large $d_h$ for high $Cc$) is a more serious problem. As indicated in Fig. 6(a) and Fig. 9(b), there exist cases in which high $Cc$ ($NCC$) such as 0.8–0.9 even when $d_h$ ($d_h^{av}$) exceeds 20. These cases complicate this screening and reduce the catalogue quality obtained by hashing-based template matching (Section 4). Fig. 18(b) presents an example of a waveform pair of false negatives in the same format as that in Fig. 18(a). In this example, despite the high $Cc$ value of 0.96 between the template and target waveforms, a large $d_h$ value of 17–20 was stably obtained, independent of the extraction window. Fig. 18(a) depicts another example between the same template and another target waveform, where a $Cc$ value of 0.98 and a small $d_h$ over 1–6 were obtained. As indicated in these figures, the proposed deep hashing model may produce false negatives owing to subtle differences in the waveform shape. Unsmoothed changes for different cutting windows shown in Fig. 18(a) and (b) also likely reflect the complicated characteristics of the obtained hash function. The percentage of false negatives must be reduced, for example, by improving training data mining and the loss function to make the deep-hash approach more effective.

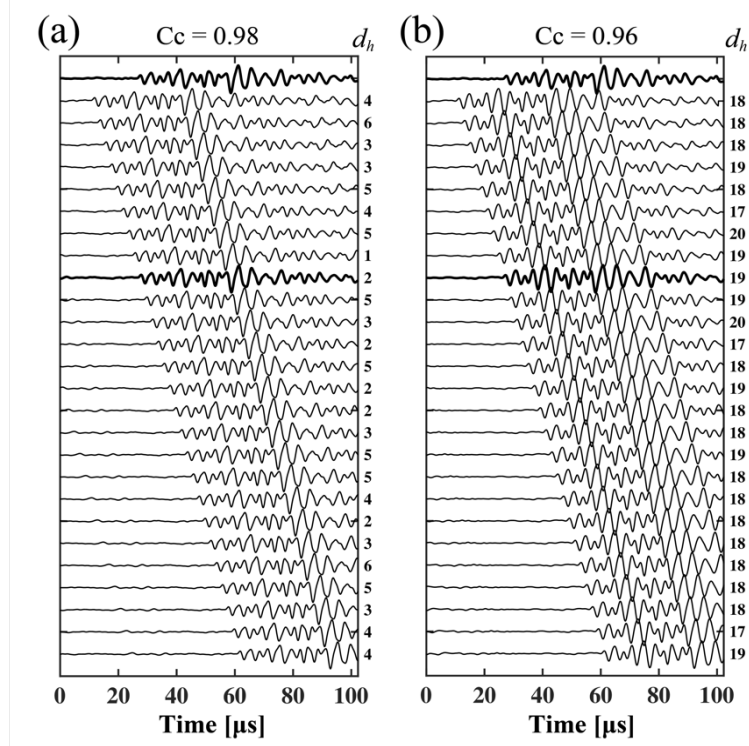

Fig. 18. Change in $d_h$ between two event waveforms. The thick line at the top indicates the template waveform, and the other thick line (10th from the top) indicates the target waveform. The $Cc$ value between them is indicated above each figure. We slid the timing of the waveform extraction window for the target event using a 2 μs step, and the corresponding $d_h$ is presented on the right of each waveform. (a) Results for waveform pairs with $Cc$ and $d_h$ values of 0.98 and 2, respectively. (b) Results for the same template and another target waveform with $Cc$ and $d_h$ values of 0.96 and 19, respectively.

### 6.3.2. Performance evaluation of cataloguing procedure incorporating deep hashing methods

The present study suggested efficient similar waveform searching based on a deep hashing technique and demonstrated that it successfully improved an AE catalogue for hydraulic fracturing experiments in the laboratory. The suggested approach is likely effective for similar waveform searching for a large-scale dataset because of smaller computation costs than the waveform cross-correlation-based approach and smaller required memory than the FAST approach, which is also based on hash functions (Section 6.2). However, for fair evaluation of these methods, we should note the difference in the assumed use cases and presumptions when using these methods.

We demonstrated the application of the deep hashing technique to the autocorrelation problem (Section 5), which is an unsupervised detection problem without template waveforms. However, our work was not completely unsupervised because labelled waveforms were used for the training of the hash function. In contrast, the FAST algorithm does not require labelled data because Hashing is performed using random substitution of binary codes, a completely data-independent method.

Therefore, the FAST algorithm can be used to detect seismic events from continuous waveform records in a completely unsupervised manner, similar to clustering-based unsupervised approaches (e.g., Seydoux et al. 2020; Steinmann et al. 2022). That is, the FAST approach is applicable to broader problems than the deep hash method in the present study. Incidentally, it is likely possible to train a deep hash function in a completely unsupervised manner using arbitrarily selected waveforms from continuous data, although the present study did not attempt to do so.

In the actual application, developing a reliable event catalogue using conventional auto-processing methods for phase picking, phase association, and location is not difficult when allowing the number of false negatives (i.e. undetected events). In fact, all training data in the present study were prepared through automatic processing by Tanaka et al. (2020). Recently, the application of deep learning methods to the catalogue developing problem has rapidly progressed, and they outperformed the conventional methods (Mousavi and Beroza 2023). Hash functions can also be trained based on deep-learning-based pipelines, possibly resulting in superior performance.

The present study successfully demonstrated that the deep hashing approach could improve the AE event catalogue. However, to prove the superiority of the method compared to some existing methods (e.g., $Cc$ or $NCC$-based template matching and auto-correlation, the FAST algorithm, and supervised classification based on deep learning) for seismic and AE cataloguing, it is necessary to evaluate the performance and computational costs across the entire analysis pipeline. This is a complicated problem because of different post-processing methods. For example, event detection using a deep learning classifier requires post-processing of the phase association process, whereas different approaches, such as double difference location, can be directly used for similar waveform search results, as we demonstrated. This different analysis procedure makes a simple comparison of performance/computing cost difficult, requiring evaluation over the entire cataloguing pipeline. For example, Scotto di Uccio et al. (2023) performed such a comparison for a combination of some methods, including FAST. They suggested that the combination of deep learning event detection/phase picking with conventional template matching is the best way to generate a high-quality catalogue. Although such a performance check is also essential for the deep hash method, we want to leave it to future studies because it is too large a problem to include in the present study.

In this study, we obtained a hash code from a single-channel AE waveform. The format of input data can be flexibly designed, and it is possible to input waveform data from multiple channels. When applying this method to seismometer records, it may be natural to hash three-component waveforms to obtain a hash code. Expanding the input data may require increasing the number of bits of the resultant hash code to maintain the capability of discrimination. This method can also be expanded for the input data different from waveforms, such as images of analogue waveforms, possibly enabling similar event searching in analogue seismic records.

# 7. CONCLUSIONS

In this study, we solved a similar waveform search problem using a deep hashing technique for the waveform data recorded by AE transducers during laboratory experiments. Our method compresses a 1024 sample waveform into 64-bit binary code, enabling a memory-efficient and low-computation-cost similarity search. Extreme dimension reduction enables the storage of information from a large-scale waveform dataset in computer memory, facilitating calculations of the Hamming distances among all hash codes, corresponding to unsupervised event detection based on waveform similarity. We calculated the distances for all combinations of $9.87 \times 10^{15}$ waveform pairs (16 channels x $35,128,124^2/2$ pairs), which has a runtime of only 15.5 h on 120 parallel threads. The deep hashing method is expected to be effective for similar waveform search problems with a large amount of continuous seismic data.

**ACKNOWLEDGEMENTS**

This research was financially supported by JSPS KAKENHI [Grant number 16H04614; 21H01191], the Kyoto University Foundation, and the Ministry of Education, Culture, Sports, Science, and Technology (MEXT) of Japan, under the Second Earthquake and Volcano Hazards Observation and Research Program (Earthquake and Volcano Hazard Reduction Research). This study used the data obtained from a research project sponsored by the Japan Oil, Gas, and Metals National Corporation. We used TensorFlow and the Keras library for all deep learning implementations.

### Data availability

The data used in the manuscript are not open to the public.

**Appendix: Efficient calculation of cross-correlation coefficient in the frequency domain**

In Section 5, we estimated the cost of the cross-correlation calculations. The benchmark calculation was conducted in the frequency domain, as presented in Algorithm 1, using openMP and cache-blocking techniques to increase the calculation speed. We simultaneously performed an inverse fast Fourier transform (FFT) for an event pair after the summation of $F\overline{F}$ among all channels, where $F$ represents the FFT of the waveform of a specific channel. In this implementation, the number of applications of inverse FFT is independent of the number of channels. This treatment corresponds to network correlation (Gibbons & Ringdal 2006), wherein a constant lag time is applied to all channel waveforms for an event pair.

> $F[k, i]$ represents the FFT result of channel $k$ record for event $i$
> for $i = 1$ to $N$
>   for $j = 1$ to $N$
>     $F'$ = summation of ($F(:, i)$ x $\overline{F(:, j)}$) along all channels
>     $f$ = inverse FFT for $F'$
>     search the maximum value of $f$ and the corresponding index
>   end
> end

Algorithm 1: Pseudocode of the cross-correlation coefficient calculation in the frequency domain.